# Resonant current in coupled inertial Brownian particles with delayed-feedback control

**Tian-Fu Gao[1], Zhi-Gang Zheng[2,*], Jin-Can Chen[3, †]**

[1]*College of Physical Science and Technology, Shenyang Normal University, Shenyang 110034, China*

[2]*College of Information Science and Engineering, Huaqiao University, Xiamen 361021, China*

[3]*Department of Physics, Xiamen University, Xiamen 361005, China*

*Corresponding authors. Email:* [*]*zgzheng@hqu.edu.cn,* [†]*jcchen@xmu.edu.cn*

The transport of a walker in rocking feedback-controlled ratchets are investigated. The walker consists of two coupled "feet" that allow the interchange of the order of the particles while the walker moves. In the underdamped case, the deterministic dynamics of the walker in a tilted asymmetric ratchet with an external periodic force is considered. It is found that the delayed feedback ratchets with a switching-on-and-off dependence of the states of the system can lead to the absolute negative mobility (ANM). In such a novel phenomenon the particles move against the bias. Moreover, the walker can acquire a series of resonant steps for different values of the current. Remarkably, it is interesting to find that the resonant current of the walker are induced by the phase locked motion that corresponds to the synchronization of the motion with the change in the frequency of the external driving. These resonant steps can be well predicted in terms of time-space symmetry analysis, which is in good agreement with dynamics simulations. The transport performances can be optimized and furthermore controlled by suitably adjusting the parameters of the delayed-feedback ratchets.

# 1 Introduction

Noise-driven transport phenomena in spatially periodic systems are a fundamental feature of many physical processes, particularly in cell biology. Mathematically speaking, these processes are often modeled in terms of Brownian particles subject to a spatially periodic potential energy environment. However, transport plays a very important role in systems that lie away from equilibrium [1]. By transport, particles can be affected by different means. The most straightforward way in inducing transport of Brownian particles is to apply a directional force on particles. Even when the applied force on average does not provide any bias to a particular direction, the asymmetry of the potential field in which the particles are moving, may bring about the passage of particles to a given direction by the rectification of thermal fluctuations. Such a phenomenon is known as the ratchet effect. This has attracted wide attention over the past two decades [2-5].

Among the many kinds of ratchets studied recently, an important class is referred to classical deterministic inertial ratchets in which the dynamics does not have any randomness or stochastic elements. Being a nonlinear system, the deterministic dynamics can be modeled as an inertial particle in a washboard potential that has been analyzed in many different situations, such as rotators [6], optical rocking ratchets [7]. If the washboard potential is acted upon by an external time-dependent driving force, it can exhibit additional interesting phenomena such as chaos [8], phase dynamics in synchronization [9], and anomalous transport [10]. Even more surprising is that the absolute negative mobility (ANM) phenomenon was observed in deterministic systems evolving in an external unbiased periodic driving and nonuniform space-dependent damping [11]. The study of inertial deterministic ratchets has acquired importance due to recent experiments on vortex and SQUID ratchets that reveal, among other things, the effect of current reversals [12].

However, ratchets can be also viewed as controllers that act on systems with the aim of inducing directed transport by breaking thermal equilibrium and certain time-space symmetries. As usual in control theory, these systems are divided into *open-loop* ratchets [2], when the actuation does not use any knowledge of the state of the system, and *closed-loop* ratchets [13,14], when information on the state of the system is used to decide how to operate on the system. These closed-loop ratchets which are also called feedback or information ratchets have recently attracted attention because of Maxwell's demon devices that are capable of maximizing the performance of ratchets [15]. In addition, experimental realizations of feedback ratchets have been recently proposed [16] and implemented [17] due to their potential relevance to nanotechnological devices. On the other hand, collective transport properties of ratchet systems and groups of interacting elements have been extensively investigated in relating to diffusions, spatiotemporal pattern dynamics and stochastic resonances.

As for directed transport, the effect of coupling among units is significant in governing the transport properties in many ratchet systems. Here, we will study the transport properties of a walker (dimer) moving through a ratchet potential. This walker comprises of two particles that open the possibility of exchanging the order when the walker walk. This model was inspired by the physics of molecular motors, in particular for kinesin, which is a motor protein that has two portions acting as feet that moves through microtubules inside cells. Further aspects of the model have been explored in Ref. [18].

In the present paper, we analyze the intriguing dynamics that emerge due to the interplay between the feedback control and the rocking. We will consider the deterministic dynamics of underdamped coupled particles moving on a tilted feedback ratchet potential that is rocked by a periodic external force and, in particular, the phenomenon of ANM [19]. Then, we show that the rocking of a feedback ratchets allows the system to improve the transport within tailored parameter regimes. The optimization of the flux performance of ratchets is potentially relevant for their nanotechnological applications, and the enhancement of the flux in flashing ratchets due to feedback has been recently verified experimentally [17]. Furthermore, the phase locked motion between the coupling and the external rocking force may lead to resonant steps of the current, which is numerically observed and theoretically predicted. In particular, the phenomenon of synchronization in ratchets and its consequences in the transport properties are discussed. Synchronization is ubiquitous in nature and is widespread in many complex systems, not only in the physical sciences but also in the life sciences. The case of phase synchronization for an *open-loop* inertial ratchet was studied previously [6,20], whereas in this paper we are considering the feedback-controlled case. We will analyze the evolution of phase space for tilted inertial coupled particles and discuss the walk mechanism of synchronization. The transport can be optimized and further controlled by adjusting the delay time and external driving. We want to emphasize that our findings are amenable to be explored in a variety of experimental realizations of Josephson Junctions or vortex ratchets, where the equations of motion is similar to our inertial deterministic ratchets. Moreover, we establish a connection between optimal transport in delayed feedback ratchets and the phenomenon of resonant steps.

## 2  The model of a walker with delayed-feedback control

Let us consider a walker with two feet and moving on an asymmetric ratchet potential $V(x)$.

The feet of the walker are indicated by the particles at coordinates $x_1$ and $x_2$, and coupled through a harmonic spring of a natural length $a$ and an elasticity constant $k$. Additionally, there is an external bias force $F$ and an fluctuating rocking force $A\cos\omega t$ with the amplitude $A$ and the frequency $\omega$. The time average of the rocking force is zero. In the absence of stochastic noise, the dynamics is exclusively deterministic.

In the following discussions we use the dimensionless notation introduced in Ref. [21]. With these assumptions, the equations of motion can be converted into its dimensionless form [22,23]. Therefore, the corresponding equations of motion for this rocked deterministic tilted ratchet in the underdamped case are given by

$$\ddot{x}_1 + \gamma\dot{x}_1 = -\beta(t)V'(x_1) + k(x_2 - x_1 - a) + A\cos\omega t + F \qquad (1)$$

and

$$\ddot{x}_2 + \gamma\dot{x}_2 = -\beta(t)V'(x_2) - k(x_2 - x_1 - a) + A\cos\omega t + F, \qquad (2)$$

where the dot and the prime denote a differential with respect to the time $t$ and the space coordinate $x$ of the Brownian particles, respectively, and the parameter $\gamma$ characterizes the friction coefficient.

Note that the control parameter $\beta(t)$ depends explicitly on the state of the walker. Therefore, this ratchet is feedback controlled, which implies an effective coupling between the particles. Furthermore, the switch of ratchet potential $V(x)$ is determined by the controller $\beta(t)$. We shall consider the delayed feedback control policy that can enhance the instant center-of-mass velocity introduced in Ref. [24]. As discussed in several studies (e.g., Refs. [13,25-27]), time delay is a rather natural phenomenon which may arise, e.g., through the finite time required for measuring or processing information from a measurement. In our delayed feedback protocol, the controller computes the force per particle due to the ratchet potential if it was on,

$$f(t) = \frac{1}{2}\sum_{i=1}^{2} F_{pot}(x_i(t)) = -\frac{1}{2}[V'(x_1(t)) + V'(x_2(t))],$$ (3)

and after a time $\tau$, switches the potential on ($\beta = 1$) if the ensemble average of the force $f(t)$ is positive or switches the potential off ($\beta = 0$) otherwise. Therefore, the delayed feedback control protocol considered is

$$\beta(t) = \begin{cases} \Theta(f(t-\tau)), & t \geq \tau \\ 0, & otherwise \end{cases},$$ (4)

where $\Theta$ is the Heaviside function. Using the real-time feedback control, the particle can be made to climb up a spiral-staircase-like potential exerted by an electric field and gains free energy larger than the amount of work done on it [28].

The two Brownian particles interact with the underlying structure of the track via the periodic asymmetric ratchet potential $V(x)$ [29],

$$V(x) = \sin\left(\frac{2\pi x}{L}\right) + \frac{\Delta}{4}\sin\left(\frac{4\pi x}{L}\right),$$ (5)

which is shown in the inset of Fig. 1, where $\Delta$ is the asymmetry parameter of the potential. Here, we set the period of the potential $L = 1$, i.e., $V(x+1) = V(x)$. However, in our feedback ratchets, the longitudinal asymmetry of the potential $V(x)$ brings about the symmetry-breaking feature of the feedback-controlled ratchets.

The most essential quantity in observing transportation processes described by the deterministic dynamics (1) and (2) is an asymptotic long-time average velocity of the center of mass $V_{cm}$, which is given by [30,31]

$$V_{cm} = \frac{1}{2} \sum_{i=1}^{2} \lim_{t \to \infty} \frac{\left[ x_i(t) - x_i(0) \right]}{t}. \tag{6}$$

When the amplitude $A = 0$, we have a tilted ratchet. In the case of $F < 0$, the tilted washboard potential is $U(x) = V(x) - Fx$, as shown in Fig. 1. When $U(x)$ is above some critical tilt, the particles slide down the fixed washboard potential and move through each period of the ratchets. Thus, the ratchet may become a rotator or self-sustained oscillator [6]. Note that the variable $x_1 - x_2$ can be positive, negative, or zero. When $x_1 - x_2 > 0$, particle $x_1$ is in the front and when $x_1 - x_2 < 0$, particle $x_2$ is in the front. Thus, the transition between the wells in the ratchet potential corresponds to an exchange of the order between the particles, as the walker alternating its two feet.

## 3   Results and discussion

The deterministic inertial ratchets of the nonequilibrium coupled system defined by Eqs. (1) and (2) exhibit a very rich and complex behavior. In order to compute the average velocity of the center of mass, we will solve numerically the equations of motion for the rocking tilted ratchet. The Runge–Kutta algorithm is used to solve the differential Eqs. (1) and (2). Therefore, we could achieve converged currents by using the position $x(0)$ randomly distributed over the periodic $L$ of the ratchet potential, with the associated initial velocities randomly chosen from an uniform distribution over the interval $[-0.2, 0.2]$.

In addition, Eqs. (1) and (2) were solved numerically and the observation time was large enough to ensure a value of the computed flux close to its asymptotic value. With the time step $h = 10^{-3}$, the total observation time was set to $5 \times 10^4$ driving periods $T_\omega = 2\pi / \omega$. For the investigation of the dynamics of our delayed feedback ratchets, we restrict the discussion here to a set of driving parameters, i.e., $\gamma = 1.546$, $L = 1$, $\Delta = 0.05$, $k = 0.01$, and $a = 0.1$.

### 3.1   Absolute negative mobility in the delayed-feedback ratchets

In Fig. 2, we depict the average velocity of the center-of-mass in the deterministic ratchets as a function of bias force $F$. The blue line shows the case without periodic driving $A\cos\omega t$, i.e., $A = 0$. Notice that the current is zero without driving amplitude until reach some critical tilt $F_C$. It can be understood that we have an asymmetric washboard potential when the amplitude $A = 0$, if the magnitude of the tilt $|F|$ is below a critical one, the walker cannot simply slide down this tilted washboard potential and therefore there is no directed transport. If the value of $|F|$ is greater than $F_C$, we can have a finite particle current that decreases monotonically with the bias force $F$.

When the periodic driving is taken into account, e.g., in the case of the amplitude $A = 2.1$, as seen in Fig. 2, remarkably, a typical example for the absolute negative mobility (ANM) phenomenon emerges, with a positive current flowing in the direction opposite to the negative bias force $F$. The ANM behavior was observed in Ref. [32] for a single particle moving in an *open-loop* zigzag-shaped potential and solely induced by noise in Refs. [19,33]. We can understand that this counterintuitive phenomenon is presumably due to the inertial effect assisted by the periodic driving force $A\cos\omega t$. In other terms, the periodic driving $A\cos\omega t$ is a prerequisite for ANM. When the bias is larger than the critical value, the deterministic dynamics is dictated by the bias and the particles display normal transport and the average velocity of the walker decreases linearly with increasing $F$.

In addition, Fig. 3 displays the time series of the two feet of the walker and the manifestation of ANM where a typical trajectory is being accelerated and the walker is climbing up the tilted ratchet potential (positive current), for the negative value of the bias $F = -0.025$, as seen in Fig. 3(a). On the contrary the normal transport of the particles in the same direction with the bias $F = -0.3$ is shown in Fig. 3(b).

## 3.2  Delay effect

Fig. 4 shows the influence of the delay time $\tau$ of the feedback ratchets on the center-of-mass average velocity $V_{cm}$ for different values of the amplitude $A$. It can be found from figure 4 that the delayed-feedback dynamics introduces complex variations in transport. For the deterministic case, only the external driving can help the particle to surmount the tilted ratchet potential $U(x)$ (see Fig. 1). Therefore, the effect of the delay time $\tau$ on the mean velocity $V_{cm}$ disappears for the amplitude $A = 0$ such that the particles cannot overcome the tilted potential in the case of the bias $F = -0.3$.

Nevertheless, it is demonstrated in Fig. 4 that $V_{cm}$ as a function of delay time $\tau$ becomes more complex for the driving amplitude $A \neq 0$. It is clearly seen that the velocity $V_{cm}$ has the maxima value at an optimal value $\tau^{opt}$ of the delay time for the amplitude $A = 2.1$. This means that the transport can be maximized by choosing the proper delay time and amplitude of the periodic driving for the present delayed feedback ratchets. However, when the driving amplitude is increased, e.g., $A = 7.6$, it is interesting to note that this current acquires a series steps for values of the current and the value of steps are rational values. Intuitively, the transport is not sensitive to the small delay time for $\tau < 0.17$, but it is the phase locked motion in the inertial ratchet and we will discuss this problem blow. Now, it is worth mentioning that the effect of the time delay is different from the previous work in which the average velocity is a monotonically decreasing function of time delay in the case of overdamped feedback ratchets [26,34]. Furthermore, it is noted that the transport can be reversed for large $\tau$. It can be understood that the action of the controller begins to be uncorrelated to the present state of the feedback ratchets and it effectively begins to act as an open-loop ratchets for large delay time.

### 3.3 Resonant current and synchronization in the delayed-feedback ratchets

In order to understand the origin of the resonant steps for current here, we show the scaled average velocity $2\pi V_{cm}/\omega$ as a function of the amplitude $A$ of the periodic driving for a fixed tilt $F = -0.3$, as shown in Fig. 5. As discussed in Fig. 4, the particles cannot jump between the wells for a very weak driving amplitudes, so the scaled average velocity is zero. However, in the case of higher amplitudes, the flux exhibits remarkable characteristic effects.

Our results show that the deterministic transport presents a steplike structure, i.e., a well-known effect for open-loop rocking ratchets (see [35] for details). The scaled current can acquire a series of clearly defined resonant steps for values of the current given by the ratio $n/m$, where $n$ and $m$ are integer numbers. This steps for current was reported for an overdamped ratchets [9,14], and for underdamped ratchets with tilt [36]. In many cases, $m = 1$ and the steps are integers, but the complicated dependence of current steps on the ac driving in the present paper can be also obtained for our delayed feedback ratchets.

Furthermore, the important to stress here is that the bias $F$ to the rocking feedback ratchets tilts the potential, so the inertial particles can slide down the tilted ratchet potential. In this way, the particles can be considered as a rotator that acts as a self-sustained oscillator, even without the external periodic forcing [6]. Consequently, we can properly synchronize the characteristic frequency $\omega_C$ of the rotator with the driving frequency $\omega$. Remember that in the case $2\pi V_{cm} = \omega_C$, a rational value of scaled average velocity $2\pi V_{cm}/\omega = n/m$ means that $\omega_C = (n/m)\omega$ for a whole range of values of the tilt. This phenomenon is called phase locked motion [8,36]. In figure 5, it should be noted that the scaled current $2\pi V_{cm}/\omega$, e.g., $2 = 2\pi V_{cm}/\omega \approx 2\pi \times 1.2/3.77$ for different delay time $\tau$ at amplitude $A = 7.6$ corresponds to a phase locked where $2\pi V_{cm}/\omega = n/m = 2/1$, so the average velocity is not sensitive to the delay time and present a platform $1.2$ in Fig. 4, and the rest resonant steps may be deduced by analogy. In fact, the resonant steps in figure 5 can be seen as the synchronization regions under the affect of periodic driving [6,20]. To elucidate the richness dynamics of information for different phase locked motion, in what follows we will make a comparison of this synchronization regions with non-synchronization region in the parameter space.

For deterministic feedback coupled ratchets, systems (1) and (2) can exhibit synchronization in the parameter space for different rational values of phase locked motion $n/m$. To describe this phenomenon in our delayed-feedback ratchets, we examine the behavior of the time dependent quantities $x_1$ and $x_2$, and the synchronization dynamics for different values of the phase locked are shown in Fig. 6. It can be seen from Fig. 6(a) that the two particles are complete synchronization, and the variable $x_1 - x_2 = 0$ for integer phase locked $n/m = 2$. However, the walker is simply synchronization for different steps of scaled average velocity in general. It is very interesting to note that the foot $x_2$ of the walker can get over the foot $x_1$ and exchange regularly in the tilted ratchet potential, as seen in Fig. 6(b). In this case, the ratchet dynamics presents the hand-over-hand mechanism for fraction phase locked, e.g., $n/m = 1/2$. Furthermore, for a negative phase locked, e.g., $n/m = -1$, the synchronization is also reached in Fig. 6(c). It can be easily seen that the foot $x_2$ is in the front all the time and the foot $x_1$ does not overcome the foot $x_2$. In this case, the ratchet dynamics presents the inchworm mechanism. Nevertheless, it turns out that if the phase locked is irrational, e.g., $n/m = 0.088495$ in Fig. 6(d), the two feet walk randomly and the variable $x_1 - x_2$ is changeable at all time, therefore in such case the coupled ratchets cannot synchronize.

The trajectories of phase space for different values of the phase locked motion $n/m$ are shown in Fig. 7, which has the same parameter space as Fig. 6. These figures also show the very rich structure in phase space, revealing the different steps of synchronization regions, as discussed above. In addition, the dynamics of the two interacting particles behaves in a similar manner with different steps, so we only show the trajectories in phase space for particle $x_1$. Notice that in the case of synchronization regions, for instance the phase locked $n/m = 2, \ 1/2$ and $-1$, as can be seen from Fig. 7(a)~(c), the trajectories in phase space are very rich structure and these phase spaces still remain periodic and obey an regular motion. However, in the region of non-synchronization, e.g., $n/m = 0.088495$, the phase space displays the irregular motion which is illustrated in Fig. 7(d).

Since the dynamics of the delayed-feedback system studied here is dissipative and deterministic, one can calculate the maximal Lyapunov exponent $\lambda_{\max}$ of the orbits in phase space in order to verify whether the transport processes for studied parameter regimes are chaotic or regular. For the regular motion in phase space shown in Fig. 7(a)~(c), we indeed find $\lambda_{\max} \approx 0$ when the exponent $\lambda_{\max}$ is computed over the same time of Eqs. (1) and (2). In this case, the perturbations on the orbit persist without exponential growth or decay, and therefore the motion of phase space is regular. However, the maximal Lyapunov exponent for the irregular motion in phase space shown in Fig. 7(d) is $\lambda_{\max} = 0.725 > 0$. Thus, the neighbouring trajectories diverge exponentially and yet remain on the same attractor, so in the non-synchron regime the motion of phase space is essentially chaotic. Therefore, it turns out that these synchronizations in the case of deterministic delayed-feedback ratchets may induced by the periodic motion of phase space.

It has been shown above that the origin of some steps (plateaus) in Fig. 5 is due to the phase locked motion between the coupled and the rocking force. We can subsequently endeavor to explore the resonant current in more detail and such steps can be predicted by the following symmetry analysis. For the rocking force $A\cos\omega t$, an important symmetry of Eqs. (1) and (2) is that for a given steady-state solution $\{x_j(t)\}$ of system, the transformation $T_{m,n}$,

$$T_{m,n}\{x_j(t)\} = \{x_{j+l}(t - 2\pi m/\omega) + 2\pi n\} \tag{7}$$

produces another steady-state solution $\{x'_j(t)\}$ [37], where $m$, $n$ are arbitrary integers. If the coupled particles are locked by the rocking force, Eqs. (1) and (2) should be invariant under the transformation $T_{m,n}$, i.e.,

$$T_{m,n}\{x_j(t)\} = \{x_j(t)\}. \tag{8}$$

This requires that the period for the phase of the ac force $A\cos\omega t$ to travel $\Theta_1 = 2\pi m$ equals that for the coupled particles shifting with a distance $\Theta_2 = Ln$ ($L$ is the period of the potential), i.e.,

$$t = \frac{\Theta_1}{\omega} = \frac{\Theta_2}{V_{cm}}, \tag{9}$$

which consequently leads to the resonant steps,

$$V_{cm} = \frac{\Theta_2}{\Theta_1} \omega = \frac{Ln}{2\pi m} \omega. \qquad (10)$$

Therefore, the resonant steps (10) can reduce to

$$V_{cm} = \frac{n}{m} \frac{L}{T_\omega}. \qquad (11)$$

In fact, a series of steps in Fig. 5 that can be well explained by Eq. (11). For the case of period $L = 1$ and integers $m$ and $n$, a series of steps can be observed for the scaled current. The phase locked motion with integers $n$ and $m$, indicates that the periodic attractor proceeds by $n$ spatial periods $L$ of the potential during $m$ periods $T_\omega$ of the external driving.

Besides the driving amplitude $A$, the frequency of the rocking force constitutes another important source intended to influence the resonant steps. The scaled average velocity of the walker as a function of the frequency $\omega$ is shown in Fig. 8. We can see the particle current that clearly exhibits the steps at the rational values of the flux. The analytical prediction of Eq. (11) agrees very well with our numerical results and the detailed structure has been reported for overdamped *open-loop* ratchets in Refs [6,20]. Notice that the resonant steps given by Eq. (11) are a series of clearly defined steps theoretically, and then, the resonant steps are induced by the collaboration between the coupling and the ac force. However, it can be observed only for a part due to the dynamics of the coupled systems in our delayed-feedback ratchets. It can be found from Fig. 8 that the presence of the coupling produces an additional dependence of the resonant steps on the spatial scales (both the period of the potential and the static length $a$). Furthermore, it indicates that the smaller the elastic coupling is (for the case of $k = 0.01$), the more the resonance step appears. However, the current scaled in this way is very large for small values of driving frequency. Mathematically, it is clear that when this driving frequency is decreased, the scaled current $2\pi V_{cm}/\omega$ increases because the center-of-mass average velocity is a finite value. When the external driving frequency increase, e.g., $\omega \to \infty$, the external driving changes very fast in this case, and the two coupled particles cannot feel the sustained driving for a short time in general. Therefore, the coupled ratchets transport only under the external bias force $F$, and the corresponding center-of-mass velocity decreases with the increase of frequency $\omega$.

# 4   Conclusion

In this paper we have studied the effects of periodic driving force on the transport of a walker in rocking feedback-controlled ratchets. The interplay between the rocking and the feedback controlled policy gives an intriguing rich dynamics. The tilted inertial ratchets can act as a rotator with a characteristic frequency, even when an external periodic driving is absent. We have analyzed the dynamics of the walker and the transport properties by calculating the average velocity (current) as a function of the parameters of the delayed feedback ratchets, such as the bias force, and the time delay, the amplitude and frequency of the driving force.

It is found that the current develops a complex structure due to the chaotic dynamics, and the remarkable phenomenon of anomalous negative mobility can be obtained in the deterministic case. It is interesting to note that for our delayed feedback control which the potential itself is switched on or off appropriately depending on the state of the coupled ratchets, the time delay is essential for the inertial ratchets and, thus, the current can be improved. In this way, we can establish a connection between the optimal transport of coupled ratchets and the closed-loop control strategy. However, the delayed feedback ratchets can acquire a series of resonant steps for different values of the current and exhibit a self-similar structure of steps, which is a typical of devil's staircase [38]. The resonant steps are owing to the phenomenon of synchronization under the affect of periodic driving, and these steps are corresponding to the phase locked motion. Within tailored parameter regimes, the phase locked induced resonant steps is observed and can be predicted theoretically. We have explored the connection in different values of the phase locked of the phase space. These facts allow us to infer that the phenomenon of synchronization might be obtained by the periodic motion of phase space.

Finally, it is worth mentioning that the results and mechanism presented here, particularly the phenomenon of ANM and resonant steps, can be well realized in experiments, including the three-junction SQUID ratchet [5,39], the rocking ratchet effect for cold atoms [40], or the Josephson vortex dynamics [4,41], where the equations of motion and the associated dynamics is similar to our nonlinear inertial feedback ratchets.

**Acknowledgments**   This work is supported by the National Natural Science Foundation of China (Grant No. 11475022, and Grant No. 11347003) and the Scientific Research Funds of Huaqiao University and the Excellent Talents Program of Shenyang Normal University (Grant No. 91400114005).

# References

1. P. K. Ghosh, P. Hänggi, F. Marchesoni, S. Martens, F. Nori, L. Schimansky-Geier, and G. Schmid, Driven Brownian transport through arrays of symmetric obstacles, *Phys. Rev. E* 85(1), 011101 (2012)

2. P. Reimann, Brownian motors: Noisy transport far from equilibrium, *Phys. Rep.* 361(2-4), 57 (2002)

3. P. K. Ghosh, V. R. Misko, F. Marchesoni, and F. Nori, Self-Propelled Janus Particles in a Ratchet: Numerical Simulations, *Phys. Rev. Lett.* 110(26), 268301 (2013)

4. J. Spiechowicz, P. Hänggi, and J. Łuczka, Josephson junction ratchet: The impact of finite capacitances, *Phys. Rev. B* 90(5), 054520 (2014)

5. J. Spiechowicz and J. Łuczka, Efficiency of the SQUID ratchet driven by external current, *New J. Phys.* 17, 023054 (2015)

6. J. L. Mateos and F. R. Alatriste, Phase synchronization in tilted inertial ratchets as chaotic rotators, *Chaos* 18(4), 043125 (2008)

7. A. V. Arzola, K. Volke-Sepúlveda, and J. L. Mateos, Dynamical analysis of an optical rocking ratchet: Theory and experiment, *Phys. Rev. E* 87(6), 062910 (2013)

8. R. L. Kautz, Noise, chaos, and the Josephson voltage standard, *Rep. Prog. Phys.* 59(8), 935 (1996)

9. F. R. Alatriste and J. L. Mateos, Phase synchronization in tilted deterministic ratchets, *Physica A* 372(2), 263 (2006)

10. D. Hennig, Current control in a tilted washboard potential via time-delayed feedback, *Phys. Rev. E* 79(4), 041114 (2009)

11. C. Mulhern, Persistence of uphill anomalous transport in inhomogeneous media, *Phys. Rev. E* 88(2), 022906 (2013)

12. C. C. de Souza Silva, J. Van de Vondel, M. Morelle, and V. V. Moshchalkov, Controlled multiple reversals of a ratchet effect, *Nature* (*London*) 440(7084), 651 (2006)

13. E. M. Craig, N. J. Kuwada, B. J. Lopez, and H. Linke, Feedback control in flashing ratchets, *Ann. Phys.* 17(2-3), 115 (2008)

14. M. Feito, J. P. Baltanás, and F. J. Cao, Rocking feedback-controlled ratchets, *Phys. Rev. E*

80(3), 031128 (2009)

15. F. J. Cao, M. Feito, and H. Touchette, Information and flux in a feedback controlled Brownian ratchet, *Physica A* 388(2-3), 113 (2009)

16. M. Feito and F. J. Cao, Time-delayed feedback control of a flashing ratchet, *Phys. Rev. E* 76(6), 061113 (2007)

17. B. J. Lopez, N. J. Kuwada, E. M. Craig, B. R. Long, and H. Linke, Realization of a Feedback Controlled Flashing Ratchet, *Phys. Rev. Lett.* 101(22), 220601 (2008)

18. M. Foroutan, Investigation of the Stochastic Dynamics of Nanomotor Protein: Effect of Bistable Potential Type, *J. Comput. Theor. Nanosci.* 6(1), 222 (2009)

19. L. Machura, M. Kostur, P. Talkner, J. Luczka, and P. Hänggi, Absolute Negative Mobility Induced by Thermal Equilibrium Fluctuations, *Phys. Rev. Lett.* 98(4), 040601 (2007)

20. J. L. Mateos and F. R. Alatriste, Phase synchronization for two Brownian motors with bistable coupling on a ratchet, *Chem. Phys.* 375(2-3), 464 (2010)

21. L. Machura, M. Kostur, P. Talkner, J. Łuczka, F. Marchesoni, and P. Hänggi, Brownian motors: Current fluctuations and rectification efficiency, *Phys. Rev. E* 70(6), 061105 (2004)

22. J. Spiechowicz, P. Hänggi, and J. Łuczka, Brownian motors in the microscale domain: Enhancement of efficiency by noise, *Phys. Rev. E* 90(3), 032104 (2014)

23. J. Spiechowicz, J. Łuczka, and L. Machura, Efficiency of transport in periodic potentials: dichotomous noise contra deterministic force, *J. Stat. Mech.*, 054038 (2016)

24. F. J. Cao, L. Dinis, and J. M. R. Parrondo, Feedback Control in a Collective Flashing Ratchet, *Phys. Rev. Lett.* 93(4), 040603 (2004)

25. M. Feito and F. J. Cao, Information and maximum power in a feedback controlled Brownian ratchet, *Eur. Phys. J. B* 59(1), 63 (2007)

26. M. Feito and F. J. Cao, Transport reversal in a delayed feedback ratchet, *Physica A* 387(18), 4553 (2008)

27. S. A. M. Loos, R. Gernert, and S. H. L. Klapp, Delay-induced transport in a rocking ratchet under feedback control, *Phys. Rev. E* 89(5), 052136 (2014)

28. S. Toyabe, T. Sagawa, M. Ueda, E. Muneyuki, and M. Sano, Experimental demonstration of information-to-energy conversion and validation of the generalized Jarzynski equality,

*Nat. Phys.* 6(12), 988 (2010)

29. B. Q. Ai and L. G. Liu, Facilitated movement of inertial Brownian motors driven by a load under an asymmetric potential, *Phys. Rev. E* 76(4), 042103 (2007)

30. B. Q. Ai, Y. F. He, and W. R. Zhong, Chirality separation of mixed chiral microswimmers in a periodic channel, *Soft Matter* 11(19), 3852 (2015)

31. B. Q. Ai, Ratchet transport powered by chiral active particles, *Sci. Rep.* 6, 18740 (2016)

32. R. Eichhorn, P. Reimann, and P. Hänggi, Paradoxical motion of a single Brownian particle: Absolute negative mobility, *Phys. Rev. E* 66(6), 066132 (2002)

33. M. Kostur, J. Łuczka, and P. Hänggi, Negative mobility induced by colored thermal fluctuations, *Phys. Rev. E* 80(5), 051121 (2009)

34. T. F. Gao, Z. G. Zheng, and J. C. Chen, Directed transport of coupled Brownian ratchets with time-delayed feedback, *Chin. Phys. B* 22(8), 080502 (2013)

35. W. T. Coffey, J. L. Déjardin, and Y. P. Kalmykov, Nonlinear noninertial response of a Brownian particle in a tilted periodic potential to a strong ac force, *Phys. Rev. E* 61(4), 4599 (2000)

36. F. R. Alatriste and J. L. Mateos, Anomalous mobility and current reversals in inertial deterministic ratchets, *Physica A* 384(2), 223 (2007)

37. Z. Zheng, M. C. Cross, and G. Hu, Collective Directed Transport of Symmetrically Coupled Lattices in Symmetric Periodic Potentials, *Phys. Rev. Lett.* 89(15), 154102 (2002)

38. C. Reichhardt and F. Nori, Phase Locking, Devil's Staircases, Farey Trees, and Arnold Tongues in Driven Vortex Lattices with Periodic Pinning, *Phys. Rev. Lett*. 82(2), 414 (1999)

39. J. Spiechowicz and J. Łuczka, Diffusion anomalies in ac-driven Brownian ratchets, *Phys. Rev. E* 91(6), 062104 (2015)

40. M. Brown and F. Renzoni, Ratchet effect in an optical lattice with biharmonic driving: A numerical analysis, *Phys. Rev. A* 77(3), 033405 (2008)

41. M. Knufinke, K. Ilin, M. Siegel, D. Koelle, R. Kleiner and E. Goldobin, Deterministic Josephson vortex ratchet with a load, *Phys. Rev. E* 85(1), 011122 (2012)

**Figure captions:**

**Fig. 1** The walker (inertial particles) climbing up a tilted ratchet potential $U(x) = V(x) - Fx$ for bias $F < 0$. The ratchet potential $V(x)$ without tilt is depicted in the inset.

**Fig. 2** The average velocity of the center of mass of the walker as a function of the bias force $F$. The blue line shows the case without the periodic driving $A\cos\omega t$ and the red line indicates the case when the periodic driving is acting on the walker with amplitude $A = 2.1$. The rest of the parameters are $\omega = 3.77$, and $\tau = 0.2$.

**Fig. 3** The trajectories for the two feet of the walker as a function of time, (a) demonstrating the counterintuitive ANM for negative bias $F = -0.025$, and (b) showing the normal transport (negative current) for bias $F = -0.3$, where the periodic driving amplitude $A = 2.1$.

**Fig. 4** The center-of-mass average velocity of the walker $V_{cm}$ varying with the time delay $\tau$ for different values of the amplitude $A$, where $\omega = 3.77$ and $F = -0.3$.

**Fig. 5** For the deterministic case without noise, the scaled average velocity $2\pi V_{cm}/\omega$ as a function of the amplitude $A$ of the periodic driving, where the tilt is $F = -0.3$, with corresponding to a frequency $\omega = 3.77$.

**Fig. 6** The trajectory of the walker for different synchronization regions, (a) frequency locking $n/m = 2$, $A = 7.6$, $\tau = 0.041$, (b) frequency locking $n/m = 1/2$, $A = 7.6$, $\tau = 0.145$, (c) frequency locking $n/m = -1$, $A = 14$, $\tau = 0.145$, and (d) non-synchronization region $n/m = 0.088495$, $A = 7.6$, $\tau = 0.175$, where $\omega = 3.77$ and $F = -0.3$.

**Fig. 7** Phase space of the walker for different synchronization regions, and the same parameter space as figure 6.

**Fig. 8** The average velocity $V_{cm}$ of the walker scaled with the frequency $\omega$ as a function of frequency $\omega$. The rest of the parameters are $A = 2.1$, $F = -0.3$, and $\tau = 0.145$.

**Fig. 1**

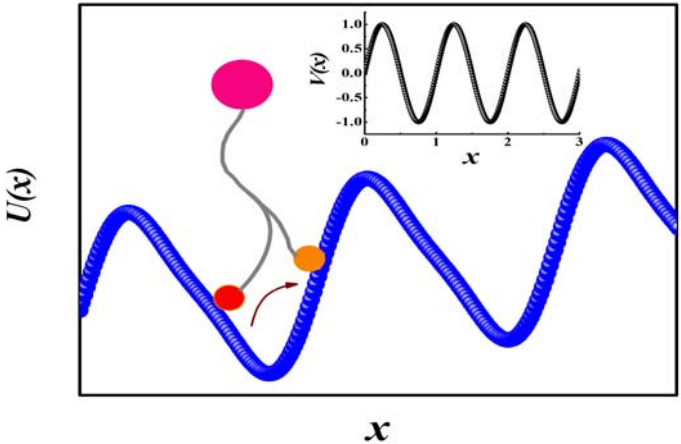

**Fig. 2**

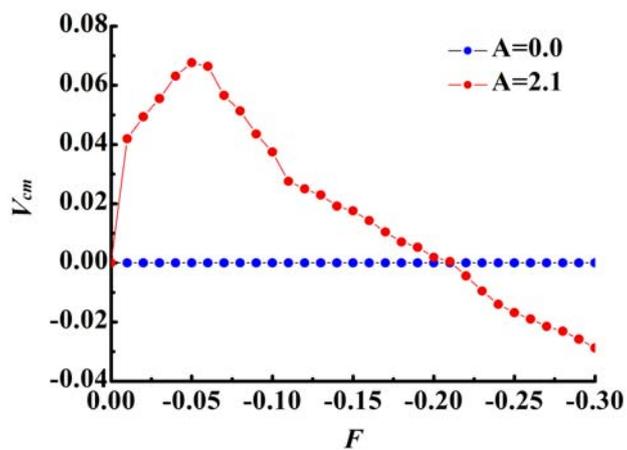

**Fig. 3**

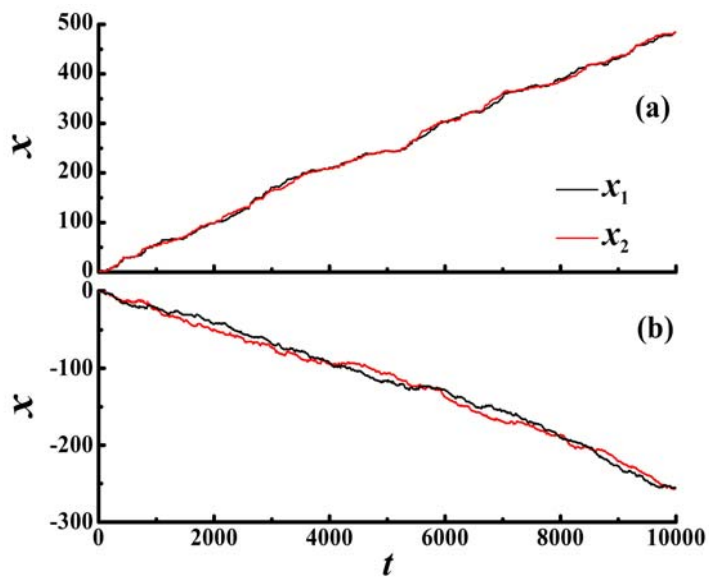

**Fig. 4**

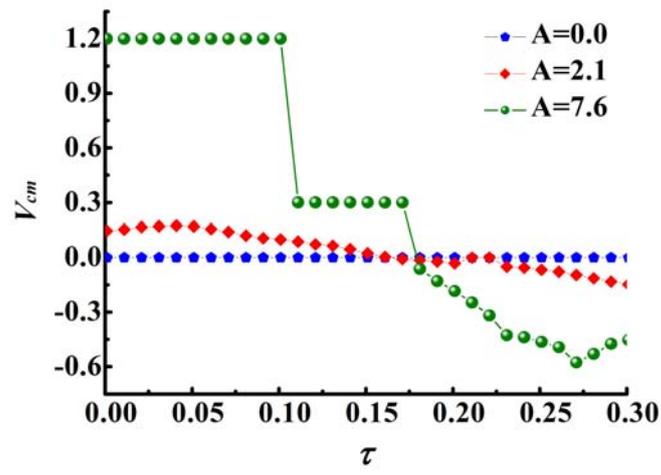

**Fig. 5**

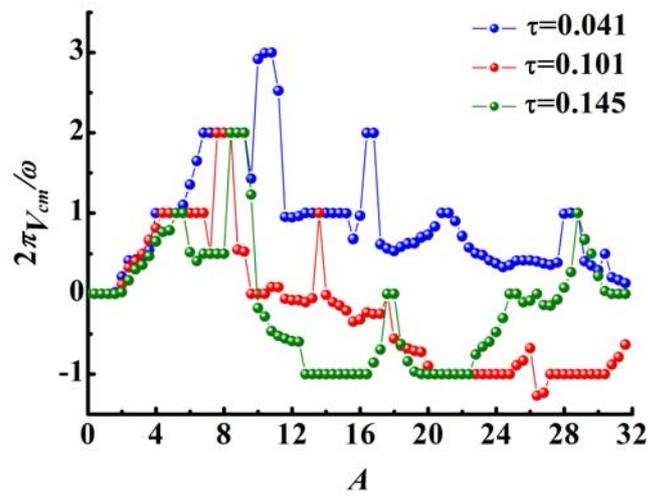

**Fig. 6**

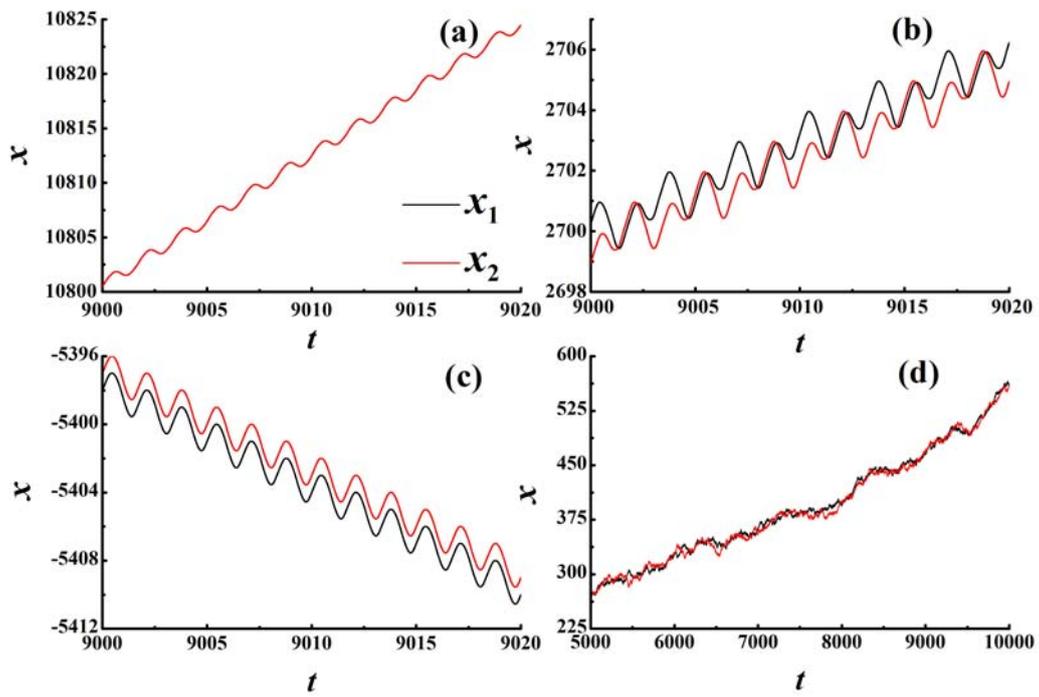

**Fig. 7**

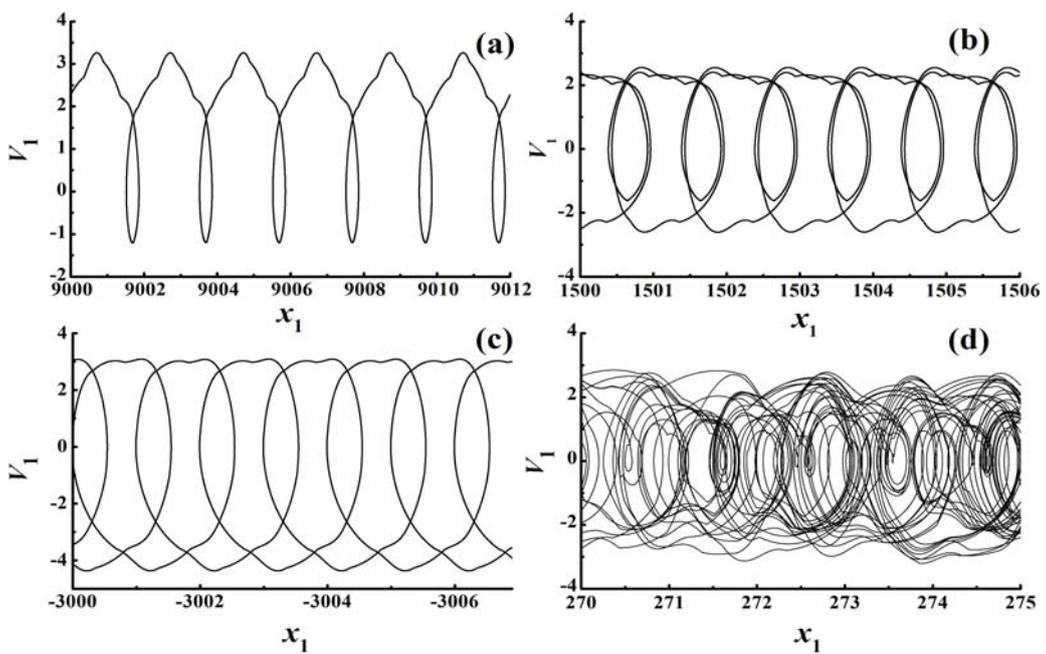

**Fig. 8**

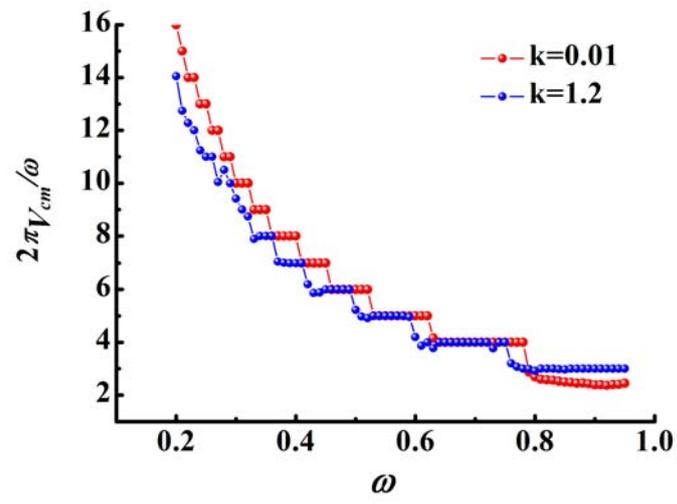